\pdfoutput=1
\documentclass[
prl,
superscriptaddress,
twocolumn,
preprintnumbers,
nofootinbib,
amsmath,amssymb,aps
]{revtex4-2}

\usepackage{graphicx}

\usepackage[
    colorlinks=true,
    linkcolor={red!50!black},
    urlcolor={green!50!black},
    citecolor={blue!80!black}
]{hyperref}

\def\mflux{\mathfrak{m}}

\usepackage{tikz}

\usetikzlibrary{
  decorations.markings,
  decorations.pathmorphing
}

\def\graphwidth{2.0}
\def\graphheight{0.8}
\def\markheight{0.05}
\def\contourgap{0.12}
\def\axescolor{black}
\def\contourcolor{red}

\tikzset{cutstyle/.style={decorate, decoration={zigzag, segment length=6, amplitude=2}, draw=black}}
\tikzset{arrow data/.style 2 args={decoration={markings, mark=at position #1 with \arrow{#2}}, postaction=decorate}}

\tikzset{
  partial ellipse/.style args={#1:#2:#3}{
      insert path={+ (#1:#3) arc (#1:#2:#3)}
  }
}

\tikzset{
  pics/torus/.style n args={3}{
    code = {
      \providecolor{pgffillcolor}{rgb}{1,1,1}
      \begin{scope}[
          yscale=cos(#3),
          outer torus/.style = {draw,line width/.expanded={\the\dimexpr2\pgflinewidth+#2*2},line join=round},
          inner torus/.style = {draw=pgffillcolor,line width={#2*2}}
        ]
        \draw[outer torus] circle(#1);\draw[inner torus] circle(#1);
        \draw[outer torus] (190:#1) arc (190:350:#1);\draw[inner torus,line cap=round] (180:#1) arc (180:360:#1);
      \end{scope}
    }
  }
}

\newcommand\drawdot[3]{
  \draw[fill] (#1, 0) circle (.7pt)
  node[black, shift=#2] {#3};
}

\newcommand\drawxmark[3]{
  \draw[-]
  (#1+\markheight,+\markheight) -- (#1-\markheight,-\markheight)
  (#1+\markheight,-\markheight) -- (#1-\markheight,+\markheight)
  (#1,0) node[black, shift=#2] {#3};
}

\newcommand\contourfigure{
  \begin{tikzpicture}[scale=2, thick]
  \draw[-latex, \axescolor] (-2*\graphwidth/3, 0) -- (+2*\graphwidth/3, 0) node[right] {$\operatorname{Re}u$};
  \draw[-latex, \axescolor] (0, -\graphheight) -- (0, +\graphheight);
  \drawdot{0}{(-120:0.5)}{$0$};
  \drawxmark{-\graphwidth/6}{(90:0.5)}{$1/\tau$};
  \draw (\graphwidth/3,0) node[red, shift=(-60:0.5)] {$R$};
  \draw[cutstyle] (0, 0) -- (+15*\graphwidth/24, 0);
  \draw[\contourcolor, thick, arrow data={0.15}{latex}, arrow data={0.25}{latex}, arrow data={0.58}{latex}, arrow data={0.75}{latex}]
      (0,\contourgap) arc (90:270:\contourgap)
      (\graphwidth/3,-\contourgap) -- (0,-\contourgap)
      (0,\contourgap) -- (\graphwidth/3,\contourgap)
      ([shift={(10:\graphwidth/3)}]0,0) arc (10:350:\graphwidth/3);
  \draw (\graphwidth/8,0) node[red, shift=(90:0.6)] {${\gamma}_{\mathrm{cut}}$};
  \draw (-\graphwidth/3,-\graphwidth/3) node[red, shift=(45:0.2)] {${\gamma}_{\mathrm{circ}}$};
  \filldraw [fill=white, draw=white] (-0.45,0.55) rectangle (-0.05,0.75);
  \draw (0, +\graphheight) node[below left] {$\operatorname{Im}u$};
  \end{tikzpicture}
}

\newcommand\WLfigure{
  \begin{tikzpicture}[scale=2, thick]
  \pic[fill=white, draw=black]{torus={1.3cm}{4mm}{65}};
  
  \draw[red, dashed] (-0.15,-0.265) [partial ellipse=-90:90:0.05 and 0.205];
  \draw[red] (-0.15,-0.27) [partial ellipse=88:272:0.05 and 0.205];
  \draw[red, -latex] (-0.2,-0.25) -- ++(0,0.01);
  
  \draw[red] (0.15,-0.27) [partial ellipse=-92:92:0.05 and 0.205];
  \draw[red, dashed] (0.15,-0.265) [partial ellipse=90:270:0.05 and 0.205];
  \draw[red, -latex] (+0.2,-0.29) -- ++(0,-0.01);
  
  \draw[thin] (0.15, -0.6) edge[<->] node[below] {$L$} (-0.15, -0.6);
  \end{tikzpicture}
}

\begin{document}

\preprint{UUITP-15/22}

\title{Exact $T\overline{T}$ deformation of two-dimensional Maxwell theory}

\author{Luca Griguolo}
 \email{luca.griguolo@unipr.it}
 \affiliation{Dipartimento SMFI, Universit\`a di Parma and INFN Gruppo Collegato di Parma,\\ Viale G.P. Usberti 7/A, 43100 Parma, Italy
}
\author{Rodolfo Panerai}
\email{rodolfo.panerai@physics.uu.se}
\affiliation{
Department of Physics and Astronomy, Uppsala University,\\ Box 516, SE-75120 Uppsala, Sweden
}
\author{Jacopo Papalini}%
 \email{jacopo.papalini@unipr.it}
\affiliation{Dipartimento SMFI, Universit\`a di Parma and INFN Gruppo Collegato di Parma,\\ Viale G.P. Usberti 7/A, 43100 Parma, Italy
}
\author{Domenico Seminara}
 \email{domenico.seminara@fi.infn.it}
\affiliation{Dipartimento di Fisica, Universit\`a di Firenze and INFN Sezione di Firenze,\\ via G. Sansone 1,
50019 Sesto Fiorentino, Italy}

\begin{abstract}
$T\overline{T}$-deformed two-dimensional quantum Maxwell theory on the torus is examined, taking into account nonperturbative effects in the deformation parameter $\mu$. We study the deformed partition function solving the relevant flow equation at the level of individual flux sectors. Summing exactly the ``instanton'' series, we obtain a well-defined expression for the partition function at arbitrary $\mu$. For $\mu>0$, the quantum spectrum of the theory experiences a truncation, the partition function reducing to a sum over a finite set of positive-energy states. For $\mu<0$ instead, the appearance of nonperturbative contributions in $\mu$ drastically modifies the structure of the partition function, regularizing its naive divergences through instanton-like subtractions. For each flux sector, we show that the semiclassical contribution is dominated by the deformed classical action. The theory is observed to undergo infinite-order phase transitions for certain values of $\mu$, associated with the vanishing of Polyakov-loop correlators.

\end{abstract}

\maketitle

\paragraph{Introduction.}\addcontentsline{toc}{section}{Introduction}
Gauge theories in low dimensions provide a valuable theoretical laboratory to test phenomena and properties likely to occur in their realistic four-dimensional counterparts. In particular, models such as Chern--Simons theory on three-manifolds \cite{Witten:1988hf} or two-dimensional Yang--Mills theory on Riemann surfaces \cite{Rusakov:1990rs, Witten:1991we, Witten:1992xu} are exactly solvable, allowing for a better understanding of quantum field theory beyond the perturbative regime. Given a model with an exact quantum solution, it is natural to look for deformations preserving (at least) part of its solvable character. Usually, one is interested in relevant deformations, which keep the theory well-defined in the UV, disregarding irrelevant ones that should, instead, uncontrollably change it. A notable counterexample is provided by a special irrelevant deformation of two-dimensional field theories, the so-called $T\overline{T}$ deformation \cite{Smirnov:2016lqw, Cavaglia:2016oda}. This preserves crucial properties of the original theory along the RG flow \cite{Smirnov:2016lqw,Cavaglia:2016oda,Baggio:2018rpv,Chang:2018dge,Jiang:2019hux,Chang:2019kiu,Aharony:2018bad} and opens the possibility to explore the dynamics of nonstandard UV fixed-points \cite{Datta:2018thy}. A striking feature of this construction is the solvability of the finite-volume spectrum $E_n$ of the deformed theory, controlled by the Burgers-type differential equation \cite{Smirnov:2016lqw, Cavaglia:2016oda}
\begin{equation}\label{EQ:Burgers}
  \frac{\partial E_n}{\partial\mu} = E_n \frac{\partial E_n}{\partial R} + \frac{P^2_n}{R},
\end{equation}
where $\mu$ is the (dimensionful) deformation, $R$ the radius of the spatial circle and $P_n$ the momentum eigenvalue.

Despite recent activity on the subject, certain aspects of $T\overline{T}$-deformed theories are still enigmatic. On the one hand, they seem intrinsically related to two-dimensional gravity \cite{Dubovsky:2017cnj,Dubovsky:2018bmo,Conti:2018tca,Ishii:2019uwk}, to random geometries \cite{Cardy:2018sdv}, and can be reformulated in terms of string theory \cite{Baggio:2018gct,Frolov:2019nrr,Hashimoto:2019wct,Sfondrini:2019smd,Callebaut:2019omt,Tolley:2019nmm}. On the other hand, they appear to encode deep aspects of holography \cite{McGough:2016lol,Giveon:2017nie,Chakraborty:2019mdf} and interesting thermodynamical properties \cite{Apolo:2019yfj,Chakraborty:2020xyz}. 
Hopes to gain further understanding of these points are generated by the fact that these deformations are special, as they preserve supersymmetry \cite{Baggio:2018rpv, Chang:2018dge, Jiang:2019hux, Chang:2019kiu}, modular invariance \cite{Aharony:2018bad}, integrability \cite{Smirnov:2016lqw, Cavaglia:2016oda}, and, in the case of Yang-Mills theory, area-preserving diffeomorphisms.

The effect of the deformation has been mainly explored starting from conformal field theories, where the $R$-dependence of the undeformed energy spectrum is fixed by conformal invariance. In this case, a radically different behavior has been observed, depending on the sign of $\mu$: positive values seem to lead to a meaningful, although non-local, theory in the UV \cite{Datta:2018thy}, while negative values are associated with a complexification of some portion of the spectrum \cite{Aharony:2018bad}. Because of this, the negative-sign deformation is believed not to make sense in the finite-volume limit. Nonperturbative effects in the parameter $\mu$ have sometimes been advocated \cite{Guica:2019nzm} to avoid this pathological behavior, but no precise computation has been performed in this direction, at least to our knowledge.

In this letter, we begin to study these problems in two-dimensional gauge theories that, as previously mentioned, have the luxury of being exactly solvable. In particular, we examine here the simplest non-trivial case: Maxwell theory on the torus.\footnote{Although placing the theory on the torus topology leads to a clear interpretation in terms of a finite-volume thermal system, the abelian theory is insensitive to the underlying topology. As such, our results apply to any genus.} We will derive the exact expression of the partition function $Z$ for both signs of the deformation parameter, solving nonperturbatively the flow equation for the $\mathrm{U}(1)$ gauge model \cite{Conti:2018jho, Ireland:2019vvj}
\begin{equation}\label{EQ:flow_equation_Z}
  \frac{\partial Z}{\partial\mu} + 2A \frac{\partial^2 Z}{\partial A^2}=0,
\end{equation}
where $A$ is the torus area.
A certain number of nontrivial features are displayed by our results that generalize to the nonabelian case \cite{future_paper}. The $T\overline{T}$ deformation of Yang--Mills theory has been already considered in \cite{Santilli:2018xux, Santilli:2020qvd, Gorsky:2020qge}, mainly in relation to the large-$N$ limit for the sphere topology. Our aim is to derive exact nonperturbative results at finite $N$.

\paragraph{Maxwell theory and its deformation.}\addcontentsline{toc}{section}{Maxwell theory and its deformation}
We consider Maxwell theory defined on the Euclidean torus $S^1\times S^1$ with lengths $R$ and $\beta$. The spectrum of the theory reads $E_n = e^2 R n^2/2$, where $e$ is the gauge coupling \cite{Manton:1985jm}. Accordingly, the partition function takes the simple form
\begin{equation}\label{EQ:Z_undeformed}
  Z = \sum_{n\in\mathbb{Z}} \mathrm{e}^{-e^2An^2/2} = \vartheta_3(\mathrm{e}^{-e^2A/2}) \;,
\end{equation}
where $A=R\beta$. A dual representation of the partition function is obtained by performing Poisson summation on \eqref{EQ:Z_undeformed},
\begin{equation}\label{EQ:Z_undeformed_m}
  Z = \sqrt{\frac{2\pi}{e^2 A}} \sum_{\mflux\in\mathbb{Z}} \mathrm{e}^{-\frac{2\pi^2}{e^2 A}\mflux^2},
\end{equation}
where the expression in the exponent represents the classical instanton action for configurations of (quantized) magnetic flux $\mflux$, while the factor in front is due to quantum fluctuations; see \cite{Cao:2013na} for technical details and relevant references. The $T\overline{T}$-deformed spectrum is readily obtained from the undeformed one by solving \eqref{EQ:Burgers}:
\begin{equation}
  E_n = \frac{e^2 R n^2/2}{1-\mu e^2 n^2} \;.
\end{equation}
Despite its simplicity, this immediately exhibits pathologies akin to the case of conformal theories. For $\mu>0$, an infinite number of energy levels become negative, namely those with $n^2>\frac{1}{\mu e^2}$, thus signaling an instability: one would like to truncate the spectrum and to explain the absence of the associated states dynamically. On the other hand, for $\mu<0$ the spectrum remains positive, but it saturates on a maximum energy $E_{\mathrm{c}} = \frac{R}{2\mu}$. One can quickly compute the density of states in this limit and observe that it diverges as $(E_{\mathrm{c}}-E)^{-3/2}$. Moreover, in both cases the naive partition function is clearly ill-defined as it is given in terms of a divergent sum.

\paragraph{Solving the flow equation in each flux sector.}\addcontentsline{toc}{section}{Solving the flow equation in each flux sector}
Since \eqref{EQ:flow_equation_Z} is linear, it should separately apply to each term of the sum in \eqref{EQ:Z_undeformed_m}. In fact, it is reasonable to expect that the deformation should lead to a well-defined result for each instanton, with the total partition function still expressible as a sum over ``deformed'' instantons and their fluctuations. Our strategy will be precisely to construct the full result as a sum over $T\overline{T}$-deformed flux sectors. As a bonus, deriving the contribution associated with each $\mflux$, we can check that the classical deformed action for $\mathrm{U}(1)$ obtained in \cite{Conti:2018tca} dominates the semiclassical expansion.

The flow equation can be easily solved by separation of variables. The generic solution,
\begin{equation}\label{EQ:flow_Z_solution}
  \frac{1}{\tau^{s}}\left[c_1 U\!\left(s,0,\frac{\alpha}{2\tau}\right) + c_2\frac{\alpha}{2\tau} {}_1F_1\!\left(s+1;2;\frac{\alpha}{2\tau}\right)\right] \;,
\end{equation}
is labeled by a real parameter $s$. Here, $c_1$ and $c_2$ are arbitrary constants, while $U$ and ${}_1F_1$ respectively denote the Tricomi and Kummer confluent hypergeometric functions. We have also introduced the two adimensional quantities $\tau = \mu e^2$ and $\alpha = e^2A$ that naturally appear in the generic solution. Denoting with ${z}_{\mflux}(\alpha,\tau)$ the deformed partition function for a given flux $\mflux$, we obtain its general form by considering a linear combination of the fundamental solutions in \eqref{EQ:flow_Z_solution}. The coefficients of such a combination are constrained to reproduce the undeformed result at $\tau=0$ and to lead to a convergent expression upon summation over $\mflux$. The behavior of \eqref{EQ:flow_Z_solution} around $\tau=0$ is sensitive to the sign of the deformation parameter. We consider the two choices separately.

The undeformed result, as it appears in \eqref{EQ:Z_undeformed_m}, can be written as a convergent power series in $1/\alpha$ with
\begin{equation}\label{EQ:z_m_undeformed}
  z_{\mflux}(\alpha,0) = \sqrt{2\pi}
  \sum_{k=0}^\infty \frac{(-2\pi^2\mflux^2)^k }{k!} \frac{1}{\alpha^{k+1/2}}.
\end{equation}
Next, we consider the expansion of \eqref{EQ:flow_Z_solution} as $\tau\to0^+$ \cite{future_paper}.
In this limit, the Kummer function blows up as $\mathrm{e}^{\frac{\alpha}{2\tau}}$ for generic values of $s$. On the other hand, $U(s,0,x) \sim x^{-s}$ for $x\to\infty$. Thus, the obvious choice to match the expansion \eqref{EQ:z_m_undeformed} is
\begin{equation}\label{EQ:z_m_positive}
  z_{\mflux}(\alpha,\tau) = {\sqrt{\frac{\pi}\tau}} \sum_{k=0}^\infty \frac{1}{k!} \bigg({-}\frac{\pi^2\mflux^2}{\tau}\bigg)^{\!k} U\!\left(k+\frac{1}{2},0,\frac{\alpha}{2\tau}\right) \,.
\end{equation}
This solution is precisely the one obtained by Borel-resumming the asymptotic series obtained through a power expansion in $\tau$ of the generic deformed flux sector $z_{\mflux}(\alpha,\tau)$ \cite{future_paper}.\footnote{Specifically, the associated Borel transform
\begin{multline}\label{EQ:Borel_transform}
  \mathcal{B}z_{\mflux}(\alpha,\zeta) = \sqrt{\frac{2\pi}{\alpha}}\sum_{k=0}^\infty \frac{1}{k!}\bigg({-}\frac{2\pi^2\mflux^2}{\alpha}\bigg)^{\!k} \\
  \times {}_2F_{1}\!\left(k+\frac{1}{2},k+\frac{3}{2};1;-\frac{2\zeta}{\alpha}\right)
\end{multline}
produces the desired result when taking a directional Laplace transform along the positive real $\zeta$-axis.}

By replacing the Tricomi function with its integral representation, we can perform the sum over $k$ and rewrite \eqref{EQ:z_m_positive} as the Fourier transform
\begin{align}\label{EQ:z_m_positive_Fourier}
  z_{\mflux}(\alpha,\tau) = \int_{-\infty}^{\infty} \mathrm{d}y \; \mathrm{e}^{2\pi\mathrm{i}\mflux y} \; \phi(y)
\end{align}
of the smooth function with compact support
\begin{align}
  \phi(y) &=
  \mathrm{e}^{-\frac{\alpha y^2}{2(1-y^2\tau)}} \, \Theta(1-y^2\tau) \;,
\end{align}
where $\Theta$ is the step function.
Now, the sum of $\mathrm{e}^{2\pi\mathrm{i}\mflux x}$ over $\mflux$ simply yields the Dirac comb of period $1$. This allows us to trivially evaluate the integral in $y$ and to obtain the full deformed partition function
\begin{align}\label{EQ:Z_maxwell_positive_tau}
  Z(\alpha,\tau)
  = \sum_{n = -\lfloor\frac{1}{\sqrt{\tau}}\rfloor}^{\lfloor\frac{1}{\sqrt{\tau}}\rfloor} \mathrm{e}^{{-}\frac{\alpha n^2}{2(1-n^2\tau)}} \;,
\end{align}
where the symbol $\lfloor x\rfloor$ stands for the integer part of $x$. In the above, we still sum over the deformed spectrum, but now all negative energies are excluded. Focusing on a specific level $n$, we see that its energy grows as $\tau$ increases and it blows up at $\tau = n^{-2}$. Above this threshold, the level drops out of the spectrum. As a consequence, only a finite number of energy levels survives when $\tau>0$. For $\tau>1$, the deformed spectrum contains only the ground state and the partition function becomes trivial: $Z(\alpha,\tau) = 1$.

When $\tau<0$, the Tricomi function in \eqref{EQ:flow_Z_solution} develops an imaginary part.
However, we can easily engineer a new ansatz for real solutions of the flow equation exploiting the second family of hypergeometrics in \eqref{EQ:flow_Z_solution},
\begin{equation}\label{EQ:z_m_negative_1}
  z_{\mflux}(\alpha, \tau)
  = \frac{\pi\alpha}{2} \sum_{k\in K} \frac{(4\pi^2\mflux^2)^{k}}{(2k)!\,(-\tau)^{k+3/2}} \, {}_1F_1\!\left(k+\frac{3}{2};2;\frac{\alpha}{2\tau}\right) \;,
\end{equation}
where $K = \{0,1,\ldots\}$.
This expression is manifestly real and reproduces the $1/\alpha$-expansion in \eqref{EQ:z_m_undeformed} for $\tau\to0^-$. In fact, ${}_1F_1(s+1,2,-x) \sim x^{-s-1}/\Gamma(1-s) + \mathrm{e}^{-x} \ldots$ for $x\to\infty$.

The presence of exponentially-suppressed terms in the expansion of the Kummer function indicates that the solution is nonanalytic at $\tau=0$. However, these are unavoidable if one wants to preserve the reality of the partition function.
Again, the origin and necessity of nonanalytic terms also emerge if we carefully examine the perturbative solution of the flow equation through the tools provided by resurgence \cite{future_paper}.\footnote{The branch cut of \eqref{EQ:Borel_transform} for $\zeta\in(-\infty,-\alpha/2)$ signals the need for instanton-like corrections in $\tau$, when $\tau<0$.}
 
\begin{figure}
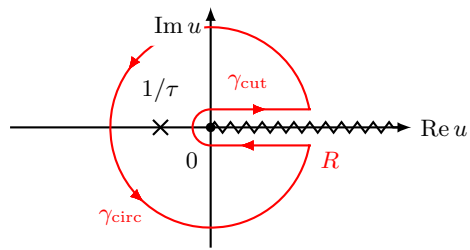

{\centering \contourfigure}
  \caption{\label{FIG:maxwell_tau_neg_contour} The contour ${\gamma}$ for the integrals in \eqref{EQ:z_m_negative_tau_integral_pre} and \eqref{EQ:z_m_negative_integral} is the union of a Hankel-like contour ${\gamma}_{\mathrm{cut}}$ and a circle ${\gamma}_{\mathrm{circle}}$ of radius $R$. In $u=1/\tau$ both integrands have an essential singularity.}
\end{figure}

Exploiting an integral representation of the Kummer function, we can perform the sum over $k$ in \eqref{EQ:z_m_negative_1} and obtain
\begin{align}\label{EQ:z_m_negative_tau_integral_pre}
  z_{\mflux}(\alpha,\tau) = - \oint_{\gamma} \mathrm{d}u \; \frac{\mathrm{i}\alpha \sinh(2\pi\mflux\sqrt{-u})\,\mathrm{e}^{-\frac{\alpha u}{2-2\tau u}}}{4\pi\mflux(\tau u-1)^2} \;,
\end{align}
where the contour $\gamma$ is depicted in FIG.~\ref{FIG:maxwell_tau_neg_contour}.
Suppose we now shrink $\gamma$ around the essential singularity in $1/\tau$ and pick up the dominant contribution at large $|\mflux|$. We can check that \eqref{EQ:z_m_negative_1} grows exponentially in this limit. Thus, the sum over the fluxes does not converge, and \eqref{EQ:z_m_negative_1} does not define a sensible partition function for $\tau<0$.

We must remark that the existence of nonanalytic contributions solving the flow equation suggests that there is a degree of arbitrariness in writing down the ansatz \eqref{EQ:z_m_negative_1}: we are free to modify it by adding any combination of solutions exponentially vanishing at $\tau=0$.
The minimal modification of \eqref{EQ:z_m_negative_1} that cancels the unsatisfactory behavior for large $|\mflux|$ is given by extending the sum to half-integers with $K = \{0,1/2,1,\ldots\}$. We will see later that this nonperturbative completion has remarkable properties.
Each new term solves the flow equation and vanishes exponentially as $\tau\to0^-$. Now 
\begin{equation}\label{EQ:z_m_negative_integral}
  z_{\mflux}(\alpha,\tau)
  = - \int_{\gamma} \mathrm{d}u \; \frac{\mathrm{i}\alpha\,\mathrm{e}^{-2\pi|\mflux|\sqrt{-u}}\,\mathrm{e}^{-\frac{\alpha u}{2-2\tau u}}}{4\pi|\mflux|(\tau u-1)^2} \;.
\end{equation}
By shrinking the contour again around the essential singularity, we can verify that $z_{\mflux}(\alpha, \tau)$ decays exponentially for large $|\mflux|$, yielding a convergent sum over the fluxes. This behavior is exactly the one suggested by the semiclassical analysis where $z_{\mflux}(\alpha, \tau)$ is expected to decay as the exponential of the deformed action \cite{Conti:2018jho}.

The easiest way to compute the integral \eqref{EQ:z_m_negative_integral} is to consider $\gamma$ as the sum of two contours $ {\gamma}_{\mathrm{cut}} \cup \gamma_{\mathrm{circle}}$ (see FIG.~\ref{FIG:maxwell_tau_neg_contour}). The integration over $\gamma_{\mathrm{circle}}$ vanishes in the limit of large $R$. Instead, the contribution of ${\gamma}_{\mathrm{cut}}$ is evaluated by taking the discontinuity of the integrand across the cut of the square root.
Upon integrating by parts and setting $u=y^2$, we find
\begin{equation}
  z_{\mflux}(\alpha,\tau)
  = 2 \int_0^{\infty} \mathrm{d}y \; \bigg(\mathrm{e}^{-\frac{\alpha}{2}\frac{y^2}{1-\tau y^2}} - \mathrm{e}^{\frac{\alpha}{2\tau}\vphantom{\frac{y^2}{1-\tau y^2}}}\bigg) \, \cos(2\pi\mflux y) \;.
\end{equation}
Once again, we observe the appearance of the Dirac comb when summing over $\mflux$, leading to
\begin{equation}\label{EQ:Z_maxwell_negative_tau}
  Z(\alpha, \tau)
  = \sum_{n = -\infty}^{\infty} \bigg(\mathrm{e}^{-\frac{\alpha}{2}\frac{n^2}{1-\tau n^2}} - \mathrm{e}^{\frac{\alpha}{2\tau}\vphantom{\frac{n^2}{1-\tau n^2}}}\bigg) \;.
\end{equation}
We notice that the entire deformed spectrum survives, but we have an additional subtraction in the partition function, nonperturbative in $\tau$, that ensures the convergence of the sum. The subtraction matches the asymptotic value of the first term for large $n$. Again, \eqref{EQ:Z_maxwell_negative_tau} solves the flow equation and reproduces the correct undeformed limit for $\tau\to 0^-$.

One would be tempted to conclude that an infinite number of nonperturbative states of energy $E_{\mathrm{c}}$ is present in the spectrum, having negative norms and regularizing the thermal trace. This feature is reminiscent of other instances of $T\overline{T}$-deformed theories which are associated with nonpositive-definite densities of states \cite{Griguolo:2021wgy}.

\paragraph{The deformed instanton action.}\addcontentsline{toc}{section}{The deformed instanton action}
In a suitable semiclassical limit, we expect that $z_{\mflux}(\alpha,\tau)$ should be dominated by the exponential of the classical deformed action evaluated on the corresponding $\mathrm{U}(1)$ instanton configuration \cite{Cavaglia:2016oda}.

First, we check this property in the case $\tau\ge0$. Performing the change of variable $y = \tau^{-1/2}\tanh(x)$ in \eqref{EQ:z_m_positive_Fourier}, we can rearrange the integral as
\begin{equation}\label{EQ:Z_m_integral_representation_for_3F2}
  {z}_{\mflux}(\alpha,\tau)
  = \frac{1}{\sqrt{\tau}} \, \int_{-\infty}^{\infty} \mathrm{d}x \; \frac{\mathrm{e}^{-4\pi^2\mflux^2\chi(x)/\alpha}}{\cosh^2(x)} \;,
\end{equation}
where 
\begin{equation}
  \chi(x) = \frac{\sinh^2(x) - 2\mathrm{i}|\mflux|\sqrt{\sigma}\tanh(x)}{2\mflux^2\sigma}
\end{equation}
and $\sigma = 4\pi^2\tau/\alpha^2$.
This representation suggests considering a double scaling in which $\alpha$ and $\tau$ are taken small with $\sigma$ fixed. In this limit, we expect the integral to be dominated by the saddles of $\chi$. Posing $x = \mathrm{i}\arctan(\sqrt{\sigma}w|\mflux|)$, the equation $\chi'(x)=0$ translates into
\begin{equation}\label{EQ:saddle_equation}
  \mflux^4 \sigma ^2 w^4 + 2 \mflux^2 \sigma w^2 - w + 1 = 0.
\end{equation}
For $\sigma<\frac{27}{256 m^2}$ we have a real saddle, which is smoothly connected to the one of the undeformed theory
\begin{equation}
w_\star = {}_3F_2\left(\frac{1}{2},\frac{3}{4},\frac{5}{4};\frac{4}{3},\frac{5}{3};\frac{256}{27}\,\mflux^2\sigma\right) \;.
\end{equation}
The integral around this saddle is easily evaluated using standard steepest descent approximation leading to
\begin{equation}\label{EQ:Z_m_double_scaling}
  z_{\mflux}(\alpha,\tau) \sim \sqrt{\frac{2\pi}{\alpha\eta}} \, \mathrm{e}^{-\frac{3\pi^2}{2\alpha\sigma}\left[{}_3F_2\left(-\frac{1}{2},-\frac{1}{4},\frac{1}{4};\frac{1}{3},\frac{2}{3};\frac{256}{27}\mflux^2\sigma\right)-1\right]} \;,
\end{equation}
where $\eta = 2w_\star^{-3/2} - w_\star^{-1} - 2\mflux^2\sigma w_\star^{1/2}$. The dominant term is exactly proportional to the exponential of the classical deformed action obtained in \cite{Cavaglia:2016oda}.

While this analysis is certainly consistent at fixed $\mathfrak{m}$, the necessity to sum over the $\mathrm{U}(1)$ fluxes poses a problem with the branch cut of the hypergeometric function, the classical action becoming complex for $\mflux^2>27\sigma/256$. A closer inspection of the interval $0<\mflux^2\sigma<27/256$ unveils a second real solution of \eqref{EQ:saddle_equation}, which provides a subdominant contribution to our expansion. The two real solutions become closer and closer, and collide exactly when $\mflux^2\sigma=27/256$, emerging further as two complex conjugate solutions. We expect them to both contribute when $\mflux^2\sigma>27/256$, combining into a real expression for the full partition function. This dramatic change in the nature of the instanton expansion is presumably related to the truncation of the spectrum observed for $\tau>0$.

We can readily repeat the same analysis in the case of $\tau<0$. At variance with the previous situation, the hypergeometric function stays real for any value of $\mflux$ (no branch cut is present when $\sigma$ is negative). The saddle point connected to the undeformed case always dominates $z_{\mflux}(\alpha,\tau)$ in the double-scaling limit. One can estimate the behavior of the instanton series at large $|\mflux|$ as
\begin{equation}\label{EQ:Z_m_double_scalingneg}
  z_{\mflux}(\alpha,\tau) \sim \mathrm{e}^{-\frac{2\pi|\mflux|}{\sqrt{-\tau}}} \;,
\end{equation}
confirming the convergence of the sum over the $\mathrm{U}(1)$ fluxes.

\paragraph{Wilson loops and quantum phase transitions.}\addcontentsline{toc}{section}{Wilson loops and quantum phase transitions}
The partition function \eqref{EQ:Z_maxwell_positive_tau} is nonanalytic whenever $\tau^{-1/2}$ is integer ($\tau = 0$ is a limit point for such a set of values). Nonetheless, it is always smooth in $\tau$. Such nonanaliticities are the signs of phase transitions of infinite order \cite{Minnhagen:1987zz}. We will show how Wilson-loop correlators \cite{Grignani:1997yg} act as order parameters for such transitions.

\begin{figure}
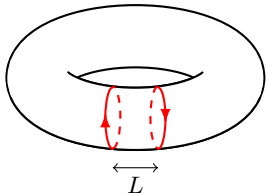

  {\centering \WLfigure}
  \caption{\label{FIG:Wilson_loops} Two parallel Wilson loops, set at a distance $L$, wrapping the fundamental cycle of radius $\beta$. In the decompactification limit, the area of the torus diverges, while $L$ and $\beta$ are kept fixed.}
\end{figure}

In accordance with our previous discussion, we introduce the partition function for an arbitrary topology with $b$ boundaries
\begin{align}
  Z_b(\alpha,\tau,\theta_1,\ldots,\theta_b) = \! \sum_{n = -\lfloor\frac{1}{\sqrt{\tau}}\rfloor}^{\lfloor\frac{1}{\sqrt{\tau}}\rfloor} \! \mathrm{e}^{{-}\frac{\alpha n^2}{2(1-n^2\tau)} + \mathrm{i}(\theta_1n_1+\ldots+\theta_1n_1)} \,,
\end{align}
where the $\theta$'s parametrize the boundary holonomies. A correlator of two homological Wilson loops is given by
\begin{align}
  \langle W_{q_1}W_{q_2}\rangle = {}& \int_0^{2\pi} \frac{\mathrm{d}\theta_1}{2\pi} \int_0^{2\pi} \frac{\mathrm{d}\theta_2}{2\pi} \; \mathrm{e}^{\mathrm{i}(\theta_1q_1+\theta_2q_2)} \cr 
  &\times \frac{Z_2(\alpha_1,\tau,\theta_1,\theta_2)\,\overline{Z_2(\alpha_2,\tau,\theta_1,\theta_2)}}{Z(\alpha_1+\alpha_2,\tau)} \;, \quad
\end{align}
where $q_1, q_2 \in \mathbb{Z}$ label the $\mathrm{U}(1)$ representations of the Wilson loops.
The above is nonvanishing for $q_1 = -q_2$ and its computation is straightforward. In the decompactification limit, where $\alpha = \alpha_1 + \alpha_2 \to \infty$ while $\alpha_2 = e^2L\beta$ is kept fixed, we find
\begin{align}
  \langle W_{q}\,W_{-q}\rangle \sim \mathrm{e}^{-e^2L\beta \frac{q^2}{2(1-\tau q^2)}} \, \Theta(1-\tau q^2) \;.
\end{align}
One can think of the two Wilson-loop insertions as the wordlines of a particle-antiparticle test pair of charge $qe$ set at a distance $L$ and wrapping around the thermal circle. See FIG.~\ref{FIG:Wilson_loops}.

For $\tau<q^{-2}$ the pair experiences an attractive potential that grows linearly with $L$, typical of a confined phase. However, as $\tau$ increases, the interaction gets stronger with an effective charge $e_{\text{eff}} = eq/\sqrt{1-\tau q^2}$. For $\tau>q^{-2}$, the potential diverges with the particles of charge $eq$ seemingly decoupling from the theory.

\emph{Conclusions and outlook.}\addcontentsline{toc}{section}{Conclusions and outlook}
We have derived the exact partition function for the $T\overline{T}$-deformed $\mathrm{U}(1)$ gauge theory on the torus. Depending on the sign of the deformation, we have found radically different behaviors.

For $\mu>0$, the spectrum of the theory undergoes a drastic reduction with only a finite number of states (namely, the ones with positive energy) surviving as indicated in Eq.~\eqref{EQ:Z_maxwell_positive_tau}. The truncation of the spectrum comes with an infinite number of quantum phase transitions, each associated with the vanishing of a certain correlator of Polyakov loops.

For $\mu<0$, the appearance of a tower of nonperturbative contributions cures the naive divergence of the partition function. Conservatively, the origin of this tower can be traced back to the existence of a state of energy $E_\mathrm{c}$ with the negative norm for each $\mathrm{U}(1)$ flux.
The spectral properties of the theory in this regime are encoded in the resolvent
\begin{align}
  R(E,\tau)
  &= \int_0^\infty \mathrm{d}\alpha \; \mathrm{e}^{\alpha E} \, Z(\alpha,\tau) \cr
  &= - \frac{\sqrt{2}\pi}{\sqrt{E}(2E\tau+1)^{3/2}} \; \cot\!\left(\frac{\sqrt{2}\pi\sqrt{E}}{\sqrt{2E\tau+1}}\right) \;. \cr
\end{align}
As already observed in the case of JT gravity \cite{Griguolo:2021wgy}, for negative values of the deformation parameter, the partition function can be reproduced by acting on the undeformed resolvent with the change of variables induced by the flow equation for the spectrum. In the case at hand, we observe that $R(E,\tau)$ is obtained from $R(E,0)$ with
\begin{align}
  E \longmapsto \frac{E}{1+2\tau E} \;.
\end{align}
The same approach cannot be used when $\mu>0$ since the above map cease to be continuous.
Still, it would be interesting to gain a better understading of \eqref{EQ:Z_maxwell_negative_tau}, in particular from the perspective of the topological composition rules of the original gauge theory \cite{Witten:1991we}. These, in fact, while fully compatible with the truncation of the spectrum for $\mu>0$, fail to naively hold for \eqref{EQ:Z_maxwell_negative_tau}.

Our construction is based on the solutions of the flow equation, an approach that can be easily extended to the nonabelian case, possibly on a general Riemann surface \cite{future_paper}. While we expect a situation similar to that of Maxwell theory for $\mu>0$, the deformed nonabelian theory should experience different behaviors for $\mu<0$, also depending on the topology. At variance with the $\mathrm{U}(1)$ case, the undeformed partition function strongly depends on the genus, and we expect further nonperturbative contributions to cure the divergencies for the sphere and the torus. In contrast, the higher-genus case could exhibit new features due to the nontrivial geometrical structures underlying the theory \cite{Witten:1991we}.

Another direction to pursue is to study the large-$N$ limit in the presence of the deformation and to explore the fate of the Gross-Taylor string expansion \cite{Gross:1993hu} (see \cite{Santilli:2018xux, Gorsky:2020qge} for results at the leading order). We expect that our techniques will enable us to approach the general case and to explore the full phase diagram of the theory on the sphere. The torus case at large $N$ should also be worth investigating since the undeformed model is closely related to conformal field theory \cite{Douglas:1993wy} and to topological strings \cite{Vafa:2004qa}.

\begin{acknowledgments}
\emph{Acknowledgments.} We thank Joseph Minahan and Stefano Negro for carefully reading the manuscript and providing suggestions. We also thank Raffaella Burioni for interesting discussions.
\end{acknowledgments}

\bibliographystyle{apsrev4-2}
\bibliography{bibliography}

\providecommand{\noopsort}[1]{}\providecommand{\singleletter}[1]{#1}%
\begin{thebibliography}{43}%
\makeatletter
\providecommand \@ifxundefined [1]{%
 \@ifx{#1\undefined}
}%
\providecommand \@ifnum [1]{%
 \ifnum #1\expandafter \@firstoftwo
 \else \expandafter \@secondoftwo
 \fi
}%
\providecommand \@ifx [1]{%
 \ifx #1\expandafter \@firstoftwo
 \else \expandafter \@secondoftwo
 \fi
}%
\providecommand \natexlab [1]{#1}%
\providecommand \enquote  [1]{``#1''}%
\providecommand \bibnamefont  [1]{#1}%
\providecommand \bibfnamefont [1]{#1}%
\providecommand \citenamefont [1]{#1}%
\providecommand \href@noop [0]{\@secondoftwo}%
\providecommand \href [0]{\begingroup \@sanitize@url \@href}%
\providecommand \@href[1]{\@@startlink{#1}\@@href}%
\providecommand \@@href[1]{\endgroup#1\@@endlink}%
\providecommand \@sanitize@url [0]{\catcode `\\12\catcode `\$12\catcode
  `\&12\catcode `\#12\catcode `\^12\catcode `\_12\catcode `\%12\relax}%
\providecommand \@@startlink[1]{}%
\providecommand \@@endlink[0]{}%
\providecommand \url  [0]{\begingroup\@sanitize@url \@url }%
\providecommand \@url [1]{\endgroup\@href {#1}{\urlprefix }}%
\providecommand \urlprefix  [0]{URL }%
\providecommand \Eprint [0]{\href }%
\providecommand \doibase [0]{https://doi.org/}%
\providecommand \selectlanguage [0]{\@gobble}%
\providecommand \bibinfo  [0]{\@secondoftwo}%
\providecommand \bibfield  [0]{\@secondoftwo}%
\providecommand \translation [1]{[#1]}%
\providecommand \BibitemOpen [0]{}%
\providecommand \bibitemStop [0]{}%
\providecommand \bibitemNoStop [0]{.\EOS\space}%
\providecommand \EOS [0]{\spacefactor3000\relax}%
\providecommand \BibitemShut  [1]{\csname bibitem#1\endcsname}%
\let\auto@bib@innerbib\@empty
\bibitem [{\citenamefont {Witten}(1989)}]{Witten:1988hf}%
  \BibitemOpen
  \bibfield  {author} {\bibinfo {author} {\bibfnamefont {E.}~\bibnamefont
  {Witten}},\ }\href {https://doi.org/10.1007/BF01217730} {\bibfield  {journal}
  {\bibinfo  {journal} {Commun. Math. Phys.}\ }\textbf {\bibinfo {volume}
  {121}},\ \bibinfo {pages} {351} (\bibinfo {year} {1989})}\BibitemShut
  {NoStop}%
\bibitem [{\citenamefont {Rusakov}(1990)}]{Rusakov:1990rs}%
  \BibitemOpen
  \bibfield  {author} {\bibinfo {author} {\bibfnamefont {B.~E.}\ \bibnamefont
  {Rusakov}},\ }\href {https://doi.org/10.1142/S0217732390000780} {\bibfield
  {journal} {\bibinfo  {journal} {Mod. Phys. Lett. A}\ }\textbf {\bibinfo
  {volume} {5}},\ \bibinfo {pages} {693} (\bibinfo {year} {1990})}\BibitemShut
  {NoStop}%
\bibitem [{\citenamefont {Witten}(1991)}]{Witten:1991we}%
  \BibitemOpen
  \bibfield  {author} {\bibinfo {author} {\bibfnamefont {E.}~\bibnamefont
  {Witten}},\ }\href {https://doi.org/10.1007/BF02100009} {\bibfield  {journal}
  {\bibinfo  {journal} {Commun. Math. Phys.}\ }\textbf {\bibinfo {volume}
  {141}},\ \bibinfo {pages} {153} (\bibinfo {year} {1991})}\BibitemShut
  {NoStop}%
\bibitem [{\citenamefont {Witten}(1992)}]{Witten:1992xu}%
  \BibitemOpen
  \bibfield  {author} {\bibinfo {author} {\bibfnamefont {E.}~\bibnamefont
  {Witten}},\ }\href {https://doi.org/10.1016/0393-0440(92)90034-X} {\bibfield
  {journal} {\bibinfo  {journal} {J. Geom. Phys.}\ }\textbf {\bibinfo {volume}
  {9}},\ \bibinfo {pages} {303} (\bibinfo {year} {1992})},\ \Eprint
  {https://arxiv.org/abs/hep-th/9204083} {arXiv:hep-th/9204083} \BibitemShut
  {NoStop}%
\bibitem [{\citenamefont {Smirnov}\ and\ \citenamefont
  {Zamolodchikov}(2017)}]{Smirnov:2016lqw}%
  \BibitemOpen
  \bibfield  {author} {\bibinfo {author} {\bibfnamefont {F.~A.}\ \bibnamefont
  {Smirnov}}\ and\ \bibinfo {author} {\bibfnamefont {A.~B.}\ \bibnamefont
  {Zamolodchikov}},\ }\href {https://doi.org/10.1016/j.nuclphysb.2016.12.014}
  {\bibfield  {journal} {\bibinfo  {journal} {Nucl. Phys. B}\ }\textbf
  {\bibinfo {volume} {915}},\ \bibinfo {pages} {363} (\bibinfo {year}
  {2017})},\ \Eprint {https://arxiv.org/abs/1608.05499} {arXiv:1608.05499
  [hep-th]} \BibitemShut {NoStop}%
\bibitem [{\citenamefont {Cavagli\`a}\ \emph {et~al.}(2016)\citenamefont
  {Cavagli\`a}, \citenamefont {Negro}, \citenamefont {Sz\'ecs\'enyi},\ and\
  \citenamefont {Tateo}}]{Cavaglia:2016oda}%
  \BibitemOpen
  \bibfield  {author} {\bibinfo {author} {\bibfnamefont {A.}~\bibnamefont
  {Cavagli\`a}}, \bibinfo {author} {\bibfnamefont {S.}~\bibnamefont {Negro}},
  \bibinfo {author} {\bibfnamefont {I.~M.}\ \bibnamefont {Sz\'ecs\'enyi}},\
  and\ \bibinfo {author} {\bibfnamefont {R.}~\bibnamefont {Tateo}},\ }\href
  {https://doi.org/10.1007/JHEP10(2016)112} {\bibfield  {journal} {\bibinfo
  {journal} {JHEP}\ }\textbf {\bibinfo {volume} {10}},\ \bibinfo {pages}
  {112}},\ \Eprint {https://arxiv.org/abs/1608.05534} {arXiv:1608.05534
  [hep-th]} \BibitemShut {NoStop}%
\bibitem [{\citenamefont {Baggio}\ \emph {et~al.}(2019)\citenamefont {Baggio},
  \citenamefont {Sfondrini}, \citenamefont {Tartaglino-Mazzucchelli},\ and\
  \citenamefont {Walsh}}]{Baggio:2018rpv}%
  \BibitemOpen
  \bibfield  {author} {\bibinfo {author} {\bibfnamefont {M.}~\bibnamefont
  {Baggio}}, \bibinfo {author} {\bibfnamefont {A.}~\bibnamefont {Sfondrini}},
  \bibinfo {author} {\bibfnamefont {G.}~\bibnamefont
  {Tartaglino-Mazzucchelli}},\ and\ \bibinfo {author} {\bibfnamefont
  {H.}~\bibnamefont {Walsh}},\ }\href {https://doi.org/10.1007/JHEP06(2019)063}
  {\bibfield  {journal} {\bibinfo  {journal} {JHEP}\ }\textbf {\bibinfo
  {volume} {06}},\ \bibinfo {pages} {063}},\ \Eprint
  {https://arxiv.org/abs/1811.00533} {arXiv:1811.00533 [hep-th]} \BibitemShut
  {NoStop}%
\bibitem [{\citenamefont {Chang}\ \emph {et~al.}(2019)\citenamefont {Chang},
  \citenamefont {Ferko},\ and\ \citenamefont {Sethi}}]{Chang:2018dge}%
  \BibitemOpen
  \bibfield  {author} {\bibinfo {author} {\bibfnamefont {C.-K.}\ \bibnamefont
  {Chang}}, \bibinfo {author} {\bibfnamefont {C.}~\bibnamefont {Ferko}},\ and\
  \bibinfo {author} {\bibfnamefont {S.}~\bibnamefont {Sethi}},\ }\href
  {https://doi.org/10.1007/JHEP04(2019)131} {\bibfield  {journal} {\bibinfo
  {journal} {JHEP}\ }\textbf {\bibinfo {volume} {04}},\ \bibinfo {pages}
  {131}},\ \Eprint {https://arxiv.org/abs/1811.01895} {arXiv:1811.01895
  [hep-th]} \BibitemShut {NoStop}%
\bibitem [{\citenamefont {Jiang}\ \emph {et~al.}(2019)\citenamefont {Jiang},
  \citenamefont {Sfondrini},\ and\ \citenamefont
  {Tartaglino-Mazzucchelli}}]{Jiang:2019hux}%
  \BibitemOpen
  \bibfield  {author} {\bibinfo {author} {\bibfnamefont {H.}~\bibnamefont
  {Jiang}}, \bibinfo {author} {\bibfnamefont {A.}~\bibnamefont {Sfondrini}},\
  and\ \bibinfo {author} {\bibfnamefont {G.}~\bibnamefont
  {Tartaglino-Mazzucchelli}},\ }\href
  {https://doi.org/10.1103/PhysRevD.100.046017} {\bibfield  {journal} {\bibinfo
   {journal} {Phys. Rev. D}\ }\textbf {\bibinfo {volume} {100}},\ \bibinfo
  {pages} {046017} (\bibinfo {year} {2019})},\ \Eprint
  {https://arxiv.org/abs/1904.04760} {arXiv:1904.04760 [hep-th]} \BibitemShut
  {NoStop}%
\bibitem [{\citenamefont {Chang}\ \emph {et~al.}(2020)\citenamefont {Chang},
  \citenamefont {Ferko}, \citenamefont {Sethi}, \citenamefont {Sfondrini},\
  and\ \citenamefont {Tartaglino-Mazzucchelli}}]{Chang:2019kiu}%
  \BibitemOpen
  \bibfield  {author} {\bibinfo {author} {\bibfnamefont {C.-K.}\ \bibnamefont
  {Chang}}, \bibinfo {author} {\bibfnamefont {C.}~\bibnamefont {Ferko}},
  \bibinfo {author} {\bibfnamefont {S.}~\bibnamefont {Sethi}}, \bibinfo
  {author} {\bibfnamefont {A.}~\bibnamefont {Sfondrini}},\ and\ \bibinfo
  {author} {\bibfnamefont {G.}~\bibnamefont {Tartaglino-Mazzucchelli}},\ }\href
  {https://doi.org/10.1103/PhysRevD.101.026008} {\bibfield  {journal} {\bibinfo
   {journal} {Phys. Rev. D}\ }\textbf {\bibinfo {volume} {101}},\ \bibinfo
  {pages} {026008} (\bibinfo {year} {2020})},\ \Eprint
  {https://arxiv.org/abs/1906.00467} {arXiv:1906.00467 [hep-th]} \BibitemShut
  {NoStop}%
\bibitem [{\citenamefont {Aharony}\ \emph {et~al.}(2019)\citenamefont
  {Aharony}, \citenamefont {Datta}, \citenamefont {Giveon}, \citenamefont
  {Jiang},\ and\ \citenamefont {Kutasov}}]{Aharony:2018bad}%
  \BibitemOpen
  \bibfield  {author} {\bibinfo {author} {\bibfnamefont {O.}~\bibnamefont
  {Aharony}}, \bibinfo {author} {\bibfnamefont {S.}~\bibnamefont {Datta}},
  \bibinfo {author} {\bibfnamefont {A.}~\bibnamefont {Giveon}}, \bibinfo
  {author} {\bibfnamefont {Y.}~\bibnamefont {Jiang}},\ and\ \bibinfo {author}
  {\bibfnamefont {D.}~\bibnamefont {Kutasov}},\ }\href
  {https://doi.org/10.1007/JHEP01(2019)086} {\bibfield  {journal} {\bibinfo
  {journal} {JHEP}\ }\textbf {\bibinfo {volume} {01}},\ \bibinfo {pages}
  {086}},\ \Eprint {https://arxiv.org/abs/1808.02492} {arXiv:1808.02492
  [hep-th]} \BibitemShut {NoStop}%
\bibitem [{\citenamefont {Datta}\ and\ \citenamefont
  {Jiang}(2018)}]{Datta:2018thy}%
  \BibitemOpen
  \bibfield  {author} {\bibinfo {author} {\bibfnamefont {S.}~\bibnamefont
  {Datta}}\ and\ \bibinfo {author} {\bibfnamefont {Y.}~\bibnamefont {Jiang}},\
  }\href {https://doi.org/10.1007/JHEP08(2018)106} {\bibfield  {journal}
  {\bibinfo  {journal} {JHEP}\ }\textbf {\bibinfo {volume} {08}},\ \bibinfo
  {pages} {106}},\ \Eprint {https://arxiv.org/abs/1806.07426} {arXiv:1806.07426
  [hep-th]} \BibitemShut {NoStop}%
\bibitem [{\citenamefont {Dubovsky}\ \emph {et~al.}(2017)\citenamefont
  {Dubovsky}, \citenamefont {Gorbenko},\ and\ \citenamefont
  {Mirbabayi}}]{Dubovsky:2017cnj}%
  \BibitemOpen
  \bibfield  {author} {\bibinfo {author} {\bibfnamefont {S.}~\bibnamefont
  {Dubovsky}}, \bibinfo {author} {\bibfnamefont {V.}~\bibnamefont {Gorbenko}},\
  and\ \bibinfo {author} {\bibfnamefont {M.}~\bibnamefont {Mirbabayi}},\ }\href
  {https://doi.org/10.1007/JHEP09(2017)136} {\bibfield  {journal} {\bibinfo
  {journal} {JHEP}\ }\textbf {\bibinfo {volume} {09}},\ \bibinfo {pages}
  {136}},\ \Eprint {https://arxiv.org/abs/1706.06604} {arXiv:1706.06604
  [hep-th]} \BibitemShut {NoStop}%
\bibitem [{\citenamefont {Dubovsky}\ \emph {et~al.}(2018)\citenamefont
  {Dubovsky}, \citenamefont {Gorbenko},\ and\ \citenamefont
  {Hern\'andez-Chifflet}}]{Dubovsky:2018bmo}%
  \BibitemOpen
  \bibfield  {author} {\bibinfo {author} {\bibfnamefont {S.}~\bibnamefont
  {Dubovsky}}, \bibinfo {author} {\bibfnamefont {V.}~\bibnamefont {Gorbenko}},\
  and\ \bibinfo {author} {\bibfnamefont {G.}~\bibnamefont
  {Hern\'andez-Chifflet}},\ }\href {https://doi.org/10.1007/JHEP09(2018)158}
  {\bibfield  {journal} {\bibinfo  {journal} {JHEP}\ }\textbf {\bibinfo
  {volume} {09}},\ \bibinfo {pages} {158}},\ \Eprint
  {https://arxiv.org/abs/1805.07386} {arXiv:1805.07386 [hep-th]} \BibitemShut
  {NoStop}%
\bibitem [{\citenamefont {Conti}\ \emph {et~al.}(2019)\citenamefont {Conti},
  \citenamefont {Negro},\ and\ \citenamefont {Tateo}}]{Conti:2018tca}%
  \BibitemOpen
  \bibfield  {author} {\bibinfo {author} {\bibfnamefont {R.}~\bibnamefont
  {Conti}}, \bibinfo {author} {\bibfnamefont {S.}~\bibnamefont {Negro}},\ and\
  \bibinfo {author} {\bibfnamefont {R.}~\bibnamefont {Tateo}},\ }\href
  {https://doi.org/10.1007/JHEP02(2019)085} {\bibfield  {journal} {\bibinfo
  {journal} {JHEP}\ }\textbf {\bibinfo {volume} {02}},\ \bibinfo {pages}
  {085}},\ \Eprint {https://arxiv.org/abs/1809.09593} {arXiv:1809.09593
  [hep-th]} \BibitemShut {NoStop}%
\bibitem [{\citenamefont {Ishii}\ \emph {et~al.}(2020)\citenamefont {Ishii},
  \citenamefont {Okumura}, \citenamefont {Sakamoto},\ and\ \citenamefont
  {Yoshida}}]{Ishii:2019uwk}%
  \BibitemOpen
  \bibfield  {author} {\bibinfo {author} {\bibfnamefont {T.}~\bibnamefont
  {Ishii}}, \bibinfo {author} {\bibfnamefont {S.}~\bibnamefont {Okumura}},
  \bibinfo {author} {\bibfnamefont {J.-I.}\ \bibnamefont {Sakamoto}},\ and\
  \bibinfo {author} {\bibfnamefont {K.}~\bibnamefont {Yoshida}},\ }\href
  {https://doi.org/10.1016/j.nuclphysb.2019.114901} {\bibfield  {journal}
  {\bibinfo  {journal} {Nucl. Phys. B}\ }\textbf {\bibinfo {volume} {951}},\
  \bibinfo {pages} {114901} (\bibinfo {year} {2020})},\ \Eprint
  {https://arxiv.org/abs/1906.03865} {arXiv:1906.03865 [hep-th]} \BibitemShut
  {NoStop}%
\bibitem [{\citenamefont {Cardy}(2018)}]{Cardy:2018sdv}%
  \BibitemOpen
  \bibfield  {author} {\bibinfo {author} {\bibfnamefont {J.}~\bibnamefont
  {Cardy}},\ }\href {https://doi.org/10.1007/JHEP10(2018)186} {\bibfield
  {journal} {\bibinfo  {journal} {JHEP}\ }\textbf {\bibinfo {volume} {10}},\
  \bibinfo {pages} {186}},\ \Eprint {https://arxiv.org/abs/1801.06895}
  {arXiv:1801.06895 [hep-th]} \BibitemShut {NoStop}%
\bibitem [{\citenamefont {Baggio}\ and\ \citenamefont
  {Sfondrini}(2018)}]{Baggio:2018gct}%
  \BibitemOpen
  \bibfield  {author} {\bibinfo {author} {\bibfnamefont {M.}~\bibnamefont
  {Baggio}}\ and\ \bibinfo {author} {\bibfnamefont {A.}~\bibnamefont
  {Sfondrini}},\ }\href {https://doi.org/10.1103/PhysRevD.98.021902} {\bibfield
   {journal} {\bibinfo  {journal} {Phys. Rev. D}\ }\textbf {\bibinfo {volume}
  {98}},\ \bibinfo {pages} {021902} (\bibinfo {year} {2018})},\ \Eprint
  {https://arxiv.org/abs/1804.01998} {arXiv:1804.01998 [hep-th]} \BibitemShut
  {NoStop}%
\bibitem [{\citenamefont {Frolov}(2020)}]{Frolov:2019nrr}%
  \BibitemOpen
  \bibfield  {author} {\bibinfo {author} {\bibfnamefont {S.}~\bibnamefont
  {Frolov}},\ }\href {https://doi.org/10.1134/S0081543820030098} {\bibfield
  {journal} {\bibinfo  {journal} {Proc. Steklov Inst. Math.}\ }\textbf
  {\bibinfo {volume} {309}},\ \bibinfo {pages} {107} (\bibinfo {year}
  {2020})},\ \Eprint {https://arxiv.org/abs/1905.07946} {arXiv:1905.07946
  [hep-th]} \BibitemShut {NoStop}%
\bibitem [{\citenamefont {Hashimoto}\ and\ \citenamefont
  {Kutasov}(2020)}]{Hashimoto:2019wct}%
  \BibitemOpen
  \bibfield  {author} {\bibinfo {author} {\bibfnamefont {A.}~\bibnamefont
  {Hashimoto}}\ and\ \bibinfo {author} {\bibfnamefont {D.}~\bibnamefont
  {Kutasov}},\ }\href {https://doi.org/10.1007/JHEP02(2020)080} {\bibfield
  {journal} {\bibinfo  {journal} {JHEP}\ }\textbf {\bibinfo {volume} {02}},\
  \bibinfo {pages} {080}},\ \Eprint {https://arxiv.org/abs/1907.07221}
  {arXiv:1907.07221 [hep-th]} \BibitemShut {NoStop}%
\bibitem [{\citenamefont {Sfondrini}\ and\ \citenamefont {van
  Tongeren}(2020)}]{Sfondrini:2019smd}%
  \BibitemOpen
  \bibfield  {author} {\bibinfo {author} {\bibfnamefont {A.}~\bibnamefont
  {Sfondrini}}\ and\ \bibinfo {author} {\bibfnamefont {S.~J.}\ \bibnamefont
  {van Tongeren}},\ }\href {https://doi.org/10.1103/PhysRevD.101.066022}
  {\bibfield  {journal} {\bibinfo  {journal} {Phys. Rev. D}\ }\textbf {\bibinfo
  {volume} {101}},\ \bibinfo {pages} {066022} (\bibinfo {year} {2020})},\
  \Eprint {https://arxiv.org/abs/1908.09299} {arXiv:1908.09299 [hep-th]}
  \BibitemShut {NoStop}%
\bibitem [{\citenamefont {Callebaut}\ \emph {et~al.}(2020)\citenamefont
  {Callebaut}, \citenamefont {Kruthoff},\ and\ \citenamefont
  {Verlinde}}]{Callebaut:2019omt}%
  \BibitemOpen
  \bibfield  {author} {\bibinfo {author} {\bibfnamefont {N.}~\bibnamefont
  {Callebaut}}, \bibinfo {author} {\bibfnamefont {J.}~\bibnamefont
  {Kruthoff}},\ and\ \bibinfo {author} {\bibfnamefont {H.}~\bibnamefont
  {Verlinde}},\ }\href {https://doi.org/10.1007/JHEP04(2020)084} {\bibfield
  {journal} {\bibinfo  {journal} {JHEP}\ }\textbf {\bibinfo {volume} {04}},\
  \bibinfo {pages} {084}},\ \Eprint {https://arxiv.org/abs/1910.13578}
  {arXiv:1910.13578 [hep-th]} \BibitemShut {NoStop}%
\bibitem [{\citenamefont {Tolley}(2020)}]{Tolley:2019nmm}%
  \BibitemOpen
  \bibfield  {author} {\bibinfo {author} {\bibfnamefont {A.~J.}\ \bibnamefont
  {Tolley}},\ }\href {https://doi.org/10.1007/JHEP06(2020)050} {\bibfield
  {journal} {\bibinfo  {journal} {JHEP}\ }\textbf {\bibinfo {volume} {06}},\
  \bibinfo {pages} {050}},\ \Eprint {https://arxiv.org/abs/1911.06142}
  {arXiv:1911.06142 [hep-th]} \BibitemShut {NoStop}%
\bibitem [{\citenamefont {McGough}\ \emph {et~al.}(2018)\citenamefont
  {McGough}, \citenamefont {Mezei},\ and\ \citenamefont
  {Verlinde}}]{McGough:2016lol}%
  \BibitemOpen
  \bibfield  {author} {\bibinfo {author} {\bibfnamefont {L.}~\bibnamefont
  {McGough}}, \bibinfo {author} {\bibfnamefont {M.}~\bibnamefont {Mezei}},\
  and\ \bibinfo {author} {\bibfnamefont {H.}~\bibnamefont {Verlinde}},\ }\href
  {https://doi.org/10.1007/JHEP04(2018)010} {\bibfield  {journal} {\bibinfo
  {journal} {JHEP}\ }\textbf {\bibinfo {volume} {04}},\ \bibinfo {pages}
  {010}},\ \Eprint {https://arxiv.org/abs/1611.03470} {arXiv:1611.03470
  [hep-th]} \BibitemShut {NoStop}%
\bibitem [{\citenamefont {Giveon}\ \emph {et~al.}(2017)\citenamefont {Giveon},
  \citenamefont {Itzhaki},\ and\ \citenamefont {Kutasov}}]{Giveon:2017nie}%
  \BibitemOpen
  \bibfield  {author} {\bibinfo {author} {\bibfnamefont {A.}~\bibnamefont
  {Giveon}}, \bibinfo {author} {\bibfnamefont {N.}~\bibnamefont {Itzhaki}},\
  and\ \bibinfo {author} {\bibfnamefont {D.}~\bibnamefont {Kutasov}},\ }\href
  {https://doi.org/10.1007/JHEP07(2017)122} {\bibfield  {journal} {\bibinfo
  {journal} {JHEP}\ }\textbf {\bibinfo {volume} {07}},\ \bibinfo {pages}
  {122}},\ \Eprint {https://arxiv.org/abs/1701.05576} {arXiv:1701.05576
  [hep-th]} \BibitemShut {NoStop}%
\bibitem [{\citenamefont {Chakraborty}\ \emph {et~al.}(2019)\citenamefont
  {Chakraborty}, \citenamefont {Giveon},\ and\ \citenamefont
  {Kutasov}}]{Chakraborty:2019mdf}%
  \BibitemOpen
  \bibfield  {author} {\bibinfo {author} {\bibfnamefont {S.}~\bibnamefont
  {Chakraborty}}, \bibinfo {author} {\bibfnamefont {A.}~\bibnamefont
  {Giveon}},\ and\ \bibinfo {author} {\bibfnamefont {D.}~\bibnamefont
  {Kutasov}},\ }\href {https://doi.org/10.1088/1751-8121/ab3710} {\bibfield
  {journal} {\bibinfo  {journal} {J. Phys. A}\ }\textbf {\bibinfo {volume}
  {52}},\ \bibinfo {pages} {384003} (\bibinfo {year} {2019})},\ \Eprint
  {https://arxiv.org/abs/1905.00051} {arXiv:1905.00051 [hep-th]} \BibitemShut
  {NoStop}%
\bibitem [{\citenamefont {Apolo}\ and\ \citenamefont
  {Song}(2020)}]{Apolo:2019yfj}%
  \BibitemOpen
  \bibfield  {author} {\bibinfo {author} {\bibfnamefont {L.}~\bibnamefont
  {Apolo}}\ and\ \bibinfo {author} {\bibfnamefont {W.}~\bibnamefont {Song}},\
  }\href {https://doi.org/10.1007/JHEP01(2020)141} {\bibfield  {journal}
  {\bibinfo  {journal} {JHEP}\ }\textbf {\bibinfo {volume} {01}},\ \bibinfo
  {pages} {141}},\ \Eprint {https://arxiv.org/abs/1907.03745} {arXiv:1907.03745
  [hep-th]} \BibitemShut {NoStop}%
\bibitem [{\citenamefont {Chakraborty}\ and\ \citenamefont
  {Hashimoto}(2020)}]{Chakraborty:2020xyz}%
  \BibitemOpen
  \bibfield  {author} {\bibinfo {author} {\bibfnamefont {S.}~\bibnamefont
  {Chakraborty}}\ and\ \bibinfo {author} {\bibfnamefont {A.}~\bibnamefont
  {Hashimoto}},\ }\href {https://doi.org/10.1007/JHEP07(2020)188} {\bibfield
  {journal} {\bibinfo  {journal} {JHEP}\ }\textbf {\bibinfo {volume} {07}},\
  \bibinfo {pages} {188}},\ \Eprint {https://arxiv.org/abs/2006.10271}
  {arXiv:2006.10271 [hep-th]} \BibitemShut {NoStop}%
\bibitem [{\citenamefont {Guica}\ and\ \citenamefont
  {Monten}(2021)}]{Guica:2019nzm}%
  \BibitemOpen
  \bibfield  {author} {\bibinfo {author} {\bibfnamefont {M.}~\bibnamefont
  {Guica}}\ and\ \bibinfo {author} {\bibfnamefont {R.}~\bibnamefont {Monten}},\
  }\href {https://doi.org/10.21468/SciPostPhys.10.2.024} {\bibfield  {journal}
  {\bibinfo  {journal} {SciPost Phys.}\ }\textbf {\bibinfo {volume} {10}},\
  \bibinfo {pages} {024} (\bibinfo {year} {2021})},\ \Eprint
  {https://arxiv.org/abs/1906.11251} {arXiv:1906.11251 [hep-th]} \BibitemShut
  {NoStop}%
\bibitem [{\citenamefont {Conti}\ \emph {et~al.}(2018)\citenamefont {Conti},
  \citenamefont {Iannella}, \citenamefont {Negro},\ and\ \citenamefont
  {Tateo}}]{Conti:2018jho}%
  \BibitemOpen
  \bibfield  {author} {\bibinfo {author} {\bibfnamefont {R.}~\bibnamefont
  {Conti}}, \bibinfo {author} {\bibfnamefont {L.}~\bibnamefont {Iannella}},
  \bibinfo {author} {\bibfnamefont {S.}~\bibnamefont {Negro}},\ and\ \bibinfo
  {author} {\bibfnamefont {R.}~\bibnamefont {Tateo}},\ }\href
  {https://doi.org/10.1007/JHEP11(2018)007} {\bibfield  {journal} {\bibinfo
  {journal} {JHEP}\ }\textbf {\bibinfo {volume} {11}},\ \bibinfo {pages}
  {007}},\ \Eprint {https://arxiv.org/abs/1806.11515} {arXiv:1806.11515
  [hep-th]} \BibitemShut {NoStop}%
\bibitem [{\citenamefont {Ireland}\ and\ \citenamefont
  {Shyam}(2020)}]{Ireland:2019vvj}%
  \BibitemOpen
  \bibfield  {author} {\bibinfo {author} {\bibfnamefont {A.}~\bibnamefont
  {Ireland}}\ and\ \bibinfo {author} {\bibfnamefont {V.}~\bibnamefont
  {Shyam}},\ }\href {https://doi.org/10.1007/JHEP07(2020)058} {\bibfield
  {journal} {\bibinfo  {journal} {JHEP}\ }\textbf {\bibinfo {volume} {07}},\
  \bibinfo {pages} {058}},\ \Eprint {https://arxiv.org/abs/1912.04686}
  {arXiv:1912.04686 [hep-th]} \BibitemShut {NoStop}%
\bibitem [{\citenamefont {Griguolo}\ \emph {et~al.}()\citenamefont {Griguolo},
  \citenamefont {Panerai}, \citenamefont {Papalini},\ and\ \citenamefont
  {Seminara}}]{future_paper}%
  \BibitemOpen
  \bibfield  {author} {\bibinfo {author} {\bibfnamefont {L.}~\bibnamefont
  {Griguolo}}, \bibinfo {author} {\bibfnamefont {R.}~\bibnamefont {Panerai}},
  \bibinfo {author} {\bibfnamefont {J.}~\bibnamefont {Papalini}},\ and\
  \bibinfo {author} {\bibfnamefont {D.}~\bibnamefont {Seminara}},\ }\href@noop
  {} {\bibinfo {title} {\textit{in preparation}}}\BibitemShut {NoStop}%
\bibitem [{\citenamefont {Santilli}\ and\ \citenamefont
  {Tierz}(2019)}]{Santilli:2018xux}%
  \BibitemOpen
  \bibfield  {author} {\bibinfo {author} {\bibfnamefont {L.}~\bibnamefont
  {Santilli}}\ and\ \bibinfo {author} {\bibfnamefont {M.}~\bibnamefont
  {Tierz}},\ }\href {https://doi.org/10.1007/JHEP01(2019)054} {\bibfield
  {journal} {\bibinfo  {journal} {JHEP}\ }\textbf {\bibinfo {volume} {01}},\
  \bibinfo {pages} {054}},\ \Eprint {https://arxiv.org/abs/1810.05404}
  {arXiv:1810.05404 [hep-th]} \BibitemShut {NoStop}%
\bibitem [{\citenamefont {Santilli}\ \emph {et~al.}(2020)\citenamefont
  {Santilli}, \citenamefont {Szabo},\ and\ \citenamefont
  {Tierz}}]{Santilli:2020qvd}%
  \BibitemOpen
  \bibfield  {author} {\bibinfo {author} {\bibfnamefont {L.}~\bibnamefont
  {Santilli}}, \bibinfo {author} {\bibfnamefont {R.~J.}\ \bibnamefont
  {Szabo}},\ and\ \bibinfo {author} {\bibfnamefont {M.}~\bibnamefont {Tierz}},\
  }\href {https://doi.org/10.1007/JHEP11(2020)086} {\bibfield  {journal}
  {\bibinfo  {journal} {JHEP}\ }\textbf {\bibinfo {volume} {11}},\ \bibinfo
  {pages} {086}},\ \Eprint {https://arxiv.org/abs/2009.00657} {arXiv:2009.00657
  [hep-th]} \BibitemShut {NoStop}%
\bibitem [{\citenamefont {Gorsky}\ \emph {et~al.}(2021)\citenamefont {Gorsky},
  \citenamefont {Pavshinkin},\ and\ \citenamefont
  {Tyutyakina}}]{Gorsky:2020qge}%
  \BibitemOpen
  \bibfield  {author} {\bibinfo {author} {\bibfnamefont {A.}~\bibnamefont
  {Gorsky}}, \bibinfo {author} {\bibfnamefont {D.}~\bibnamefont {Pavshinkin}},\
  and\ \bibinfo {author} {\bibfnamefont {A.}~\bibnamefont {Tyutyakina}},\
  }\href {https://doi.org/10.1007/JHEP03(2021)142} {\bibfield  {journal}
  {\bibinfo  {journal} {JHEP}\ }\textbf {\bibinfo {volume} {03}},\ \bibinfo
  {pages} {142}},\ \Eprint {https://arxiv.org/abs/2012.09467} {arXiv:2012.09467
  [hep-th]} \BibitemShut {NoStop}%
\bibitem [{\citenamefont {Manton}(1985)}]{Manton:1985jm}%
  \BibitemOpen
  \bibfield  {author} {\bibinfo {author} {\bibfnamefont {N.~S.}\ \bibnamefont
  {Manton}},\ }\href {https://doi.org/10.1016/0003-4916(85)90199-X} {\bibfield
  {journal} {\bibinfo  {journal} {Annals Phys.}\ }\textbf {\bibinfo {volume}
  {159}},\ \bibinfo {pages} {220} (\bibinfo {year} {1985})}\BibitemShut
  {NoStop}%
\bibitem [{\citenamefont {Cao}\ \emph {et~al.}(2013)\citenamefont {Cao},
  \citenamefont {van Caspel},\ and\ \citenamefont {Zhitnitsky}}]{Cao:2013na}%
  \BibitemOpen
  \bibfield  {author} {\bibinfo {author} {\bibfnamefont {C.}~\bibnamefont
  {Cao}}, \bibinfo {author} {\bibfnamefont {M.}~\bibnamefont {van Caspel}},\
  and\ \bibinfo {author} {\bibfnamefont {A.~R.}\ \bibnamefont {Zhitnitsky}},\
  }\href {https://doi.org/10.1103/PhysRevD.87.105012} {\bibfield  {journal}
  {\bibinfo  {journal} {Phys. Rev. D}\ }\textbf {\bibinfo {volume} {87}},\
  \bibinfo {pages} {105012} (\bibinfo {year} {2013})},\ \Eprint
  {https://arxiv.org/abs/1301.1706} {arXiv:1301.1706 [hep-th]} \BibitemShut
  {NoStop}%
\bibitem [{\citenamefont {Griguolo}\ \emph {et~al.}(2022)\citenamefont
  {Griguolo}, \citenamefont {Panerai}, \citenamefont {Papalini},\ and\
  \citenamefont {Seminara}}]{Griguolo:2021wgy}%
  \BibitemOpen
  \bibfield  {author} {\bibinfo {author} {\bibfnamefont {L.}~\bibnamefont
  {Griguolo}}, \bibinfo {author} {\bibfnamefont {R.}~\bibnamefont {Panerai}},
  \bibinfo {author} {\bibfnamefont {J.}~\bibnamefont {Papalini}},\ and\
  \bibinfo {author} {\bibfnamefont {D.}~\bibnamefont {Seminara}},\ }\href
  {https://doi.org/10.1103/PhysRevD.105.046015} {\bibfield  {journal} {\bibinfo
   {journal} {Phys. Rev. D}\ }\textbf {\bibinfo {volume} {105}},\ \bibinfo
  {pages} {046015} (\bibinfo {year} {2022})},\ \Eprint
  {https://arxiv.org/abs/2106.01375} {arXiv:2106.01375 [hep-th]} \BibitemShut
  {NoStop}%
\bibitem [{\citenamefont {Minnhagen}(1987)}]{Minnhagen:1987zz}%
  \BibitemOpen
  \bibfield  {author} {\bibinfo {author} {\bibfnamefont {P.}~\bibnamefont
  {Minnhagen}},\ }\href {https://doi.org/10.1103/RevModPhys.59.1001} {\bibfield
   {journal} {\bibinfo  {journal} {Rev. Mod. Phys.}\ }\textbf {\bibinfo
  {volume} {59}},\ \bibinfo {pages} {1001} (\bibinfo {year}
  {1987})}\BibitemShut {NoStop}%
\bibitem [{\citenamefont {Grignani}\ \emph {et~al.}(1997)\citenamefont
  {Grignani}, \citenamefont {Paniak}, \citenamefont {Semenoff},\ and\
  \citenamefont {Sodano}}]{Grignani:1997yg}%
  \BibitemOpen
  \bibfield  {author} {\bibinfo {author} {\bibfnamefont {G.}~\bibnamefont
  {Grignani}}, \bibinfo {author} {\bibfnamefont {L.}~\bibnamefont {Paniak}},
  \bibinfo {author} {\bibfnamefont {G.~W.}\ \bibnamefont {Semenoff}},\ and\
  \bibinfo {author} {\bibfnamefont {P.}~\bibnamefont {Sodano}},\ }\href
  {https://doi.org/10.1006/aphy.1997.5722} {\bibfield  {journal} {\bibinfo
  {journal} {Annals Phys.}\ }\textbf {\bibinfo {volume} {260}},\ \bibinfo
  {pages} {275} (\bibinfo {year} {1997})},\ \Eprint
  {https://arxiv.org/abs/hep-th/9705102} {arXiv:hep-th/9705102} \BibitemShut
  {NoStop}%
\bibitem [{\citenamefont {Gross}\ and\ \citenamefont
  {Taylor}(1993)}]{Gross:1993hu}%
  \BibitemOpen
  \bibfield  {author} {\bibinfo {author} {\bibfnamefont {D.~J.}\ \bibnamefont
  {Gross}}\ and\ \bibinfo {author} {\bibfnamefont {W.}~\bibnamefont {Taylor}},\
  }\href {https://doi.org/10.1016/0550-3213(93)90403-C} {\bibfield  {journal}
  {\bibinfo  {journal} {Nucl. Phys. B}\ }\textbf {\bibinfo {volume} {400}},\
  \bibinfo {pages} {181} (\bibinfo {year} {1993})},\ \Eprint
  {https://arxiv.org/abs/hep-th/9301068} {arXiv:hep-th/9301068} \BibitemShut
  {NoStop}%
\bibitem [{\citenamefont {Douglas}(1993)}]{Douglas:1993wy}%
  \BibitemOpen
  \bibfield  {author} {\bibinfo {author} {\bibfnamefont {M.~R.}\ \bibnamefont
  {Douglas}},\ }in\ \href@noop {} {\emph {\bibinfo {booktitle} {{NATO Advanced
  Research Workshop on New Developments in String Theory, Conformal Models and
  Topological Field Theory}}}}\ (\bibinfo {year} {1993})\ \Eprint
  {https://arxiv.org/abs/hep-th/9311130} {arXiv:hep-th/9311130} \BibitemShut
  {NoStop}%
\bibitem [{\citenamefont {Vafa}(2004)}]{Vafa:2004qa}%
  \BibitemOpen
  \bibfield  {author} {\bibinfo {author} {\bibfnamefont {C.}~\bibnamefont
  {Vafa}},\ }\href@noop {} {\bibinfo {title} {{Two dimensional Yang-Mills,
  black holes and topological strings}}} (\bibinfo {year} {2004}),\ \Eprint
  {https://arxiv.org/abs/hep-th/0406058} {arXiv:hep-th/0406058} \BibitemShut
  {NoStop}%
\end{thebibliography}%

\end{document}